\documentclass[11pt,epsfig]{article}
\usepackage{graphicx,amsmath,amssymb}

\newlength{\dinwidth}
\newlength{\dinmargin}
\def\lapproxeq{\lower .7ex\hbox{$\;\stackrel{\textstyle
<}{\sim}\;$}}
\def\gapproxeq{\lower .7ex\hbox{$\;\stackrel{\textstyle
>}{\sim}\;$}}
\def\gtrsim{\lower .7ex\hbox{$\;\stackrel{\textstyle
>}{\sim}\;$}}
\def\lesim{\lower .7ex\hbox{$\;\stackrel{\textstyle
<}{\sim}\;$}}
\def\bb{b\bar{b}}

\begin{document}
\begin{flushright}
IPPP/06/32 \\
DCPT/06/64 \\
16th May 2006 \\

\end{flushright}

\vspace*{0.5cm}

\begin{center}
{\Large \bf Diffractive Higgs production at the LHC\footnote{Presented at the XIV International Workshop
on Deep Inelastic Scattering, DIS 2006, 20-24 April 2006, Tsukuba, Japan}}

\vspace*{1cm}
\textsc{A.D. Martin$^a$, V.A.~Khoze$^{a,b}$ and M.G. Ryskin$^{a,b}$} \\

\vspace*{0.5cm}
$^a$ Department of Physics and Institute for
Particle Physics Phenomenology, \\
University of Durham, DH1 3LE, UK \\
$^b$ Petersburg Nuclear Physics Institute, Gatchina,
St.~Petersburg, 188300, Russia \\

\end{center}

\vspace*{0.5cm}
\begin{abstract}
We review recent issues concerning exclusive Higgs production at the LHC, and related processes at the Tevatron and HERA.
\end{abstract}

\vspace*{0.5cm}

\begin{figure}
\begin{center}
\includegraphics[height=8cm]{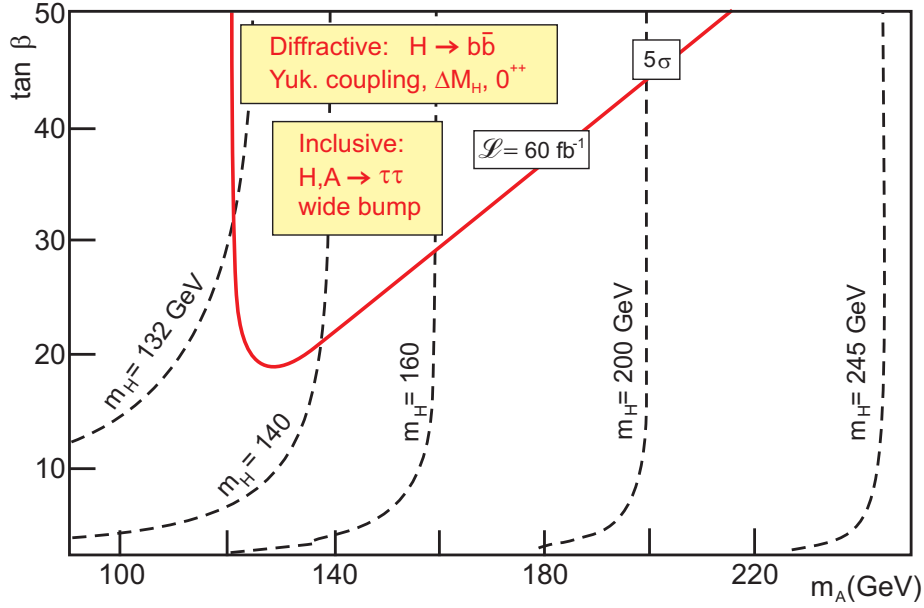}
\caption{Adapted from a preliminary plot due to Tasevsky et al. It shows the region in the ($m_A, {\rm tan}\beta$) plane with more than a $5\sigma$ signal for the production of the heavy $0^{++}$ Higgs boson by the exclusive process $pp \to p+(H \to \bb)+p$ for a $60 ~{\rm fb}^{-1}$ LHC luminosity.  Moreover, such events would determine $B(H \to \bb)\sigma$ and an accurate value of $m_H$.  This is in contrast to the broad $H,A \to \tau\tau$ conventional signal in this ($m_A, {\rm tan}\beta$) domain.} 
\label{fig:H}
\end{center}
\end{figure}

In spite of the small predicted cross section the exclusive Higgs production,
$pp \to p+H+p$, has the following advantages:
\begin{itemize}
\item[(a)]
The mass of the Higgs boson can be measured
with high accuracy (with mass resolution $\sigma(M)\sim 1$ GeV) by measuring the
missing mass to the forward outgoing protons, {\it provided} that they can be accurately tagged far away (420m) from the interaction point\cite{FP420}. 
\item[(b)]
The leading order $b\bar b$
  QCD background is suppressed by the P-even $J_z=0$ selection rule, and by colour and spin factors\cite{KMRmm}. It should be possible to achieve a signal-to-background ratio of $\sim 1$ for SM Higgs of mass $m \lapproxeq 140$ GeV. For an 
LHC luminosity of $\sim 60 ~{\rm fb}^{-1}$ we expect $\sim 12$ observable events, {\it after} accounting for signal efficiencies and various cuts\footnote{Early estimates of the signal-to-background ratio can be found in De Roeck et al.\cite{DKMOR}}.
\item[(c)]
There is a very clean environment for the
exclusive process.
\item[(d)]
Extending the study to SUSY Higgs bosons, there are regions of SUSY parameter space where the conventional inclusive Higgs search modes are suppressed, whereas the
exclusive diffractive $0^{++}$ signal is
enhanced, and even such that both the $h$ and $H$ $0^{++}$ bosons may be detected\cite{KKMRext}; an example is shown in Fig.~$\ref{fig:H}$    
\end{itemize}
\begin{figure}
\begin{center}
\includegraphics[height=4.5cm]{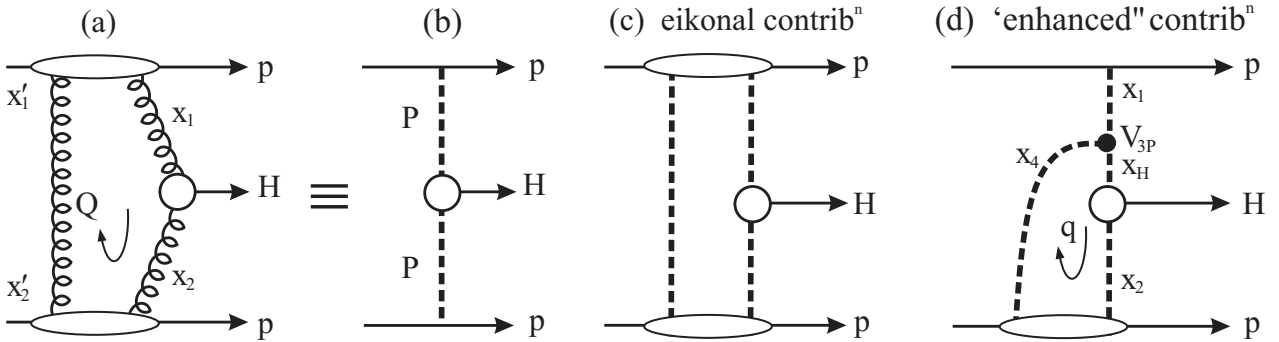}
\caption[Schematic]{Schematic diagrams for exclusive Higgs production,
$pp \to p+H+p$. The survival factor, ${\hat S}^2$, is ignored in (a,b). The {\it eikonal} diagrams, like (c), give ${\hat S}^2\simeq 0.026$ at the LHC\cite{KKMRext,lonnblad,gotsman,BM},
whereas we argue\cite{KMRenh} that {\it enhanced} diagrams, like (d), do not give a significant contribution.}
\label{fig:D}
\end{center}
\end{figure}

The basic mechanism for the exclusive process, $pp\to p+H+p$, is
shown in Fig.~$\ref{fig:D}$(a,b).  The $t$-integrated cross section is of the form\cite{KMR,INC}
\begin{equation}
\sigma \sim \frac{{\hat S}^2}{b^2} \left| N\int\frac{dQ^2_t}{Q^4_t}\: f_g(x_1, x_1', Q_t^2, \mu^2)f_g(x_2,x_2',Q_t^2,\mu^2)~ \right| ^2, 
\label{eq:M}
\end{equation}
where $b/2$ is the $t$-slope of the proton-Pomeron vertex, and the constant $N$ is known in terms of the $H\to gg$ decay width. The probability amplitudes, $f_g$, to find the appropriate pairs of
$t$-channel gluons ($x_1,x_1'$) and ($x_2,x_2'$), are given by the skewed unintegrated gluon densities at a {\it hard} scale $\mu \sim m_H/2$. The $x$ are sufficiently small that the $f_g$'s are given in terms of the conventional gluon density $g(x,Q_t^2)$, together with a known Sudakov factor which ensures the integral over $Q_t$ is infrared stable.
Thus the amplitude-squared factor, $|M_0|^2$ in (\ref{eq:M}), may
be calculated using pQCD. The factor, ${\hat S}^2 $, is the probability that the rapidity gaps survive against population by secondary hadrons. The {\it eikonal} diagrams, like (c), give ${\hat S}^2\simeq 0.026$ at the LHC, which results in $\sigma_{\rm excl}(H) \simeq 3$ fb for a SM Higgs.

Besides the soft eikonal rescattering, 
there is the possibility that the rapidity gaps may be filled by the
secondaries created in the rescattering of the intermediate
partons. Formally this effect is described by the semi-enhanced (and/or
enhanced) reggeon diagrams. One such diagram is shown schematically 
in Fig.~$\ref{fig:D}$(d). Since the intermediate gluons have a
relatively large transverse momenta, there has been an attempt\cite{BM} to calculate such a contribution using pQCD.  They evaluated an amplitude
of the form\footnote{Here the triple-Pomeron vertex $V_{3P}$ is defined slightly differently to that
in ref.\cite{BM}.}
\begin{equation}
M_1~\sim~\int \frac{dx_4}{x_4} \int\frac{d^2 q_t}{2\pi^2} \int\frac{d^2 k_{t,4}}{k^4_{t,4}}~f_g(x_4,k^2_{t,4},...)
~V_{3P}~M_0,
\label{eq:enh}
\end{equation}
where the unintegrated gluon density $f_g(x_4,...)$ was calculated using 
the Balitsky-Kovchegov equation, and where the leading log expression was used
for the QCD triple-Pomeron vertex, $V_{3P}$. In this way they found that the enhanced diagrams might give a rather large (negative) correction to exclusive Higgs boson production at the LHC energy, leading to a significantly reduced exclusive Higgs signal. The infrared divergency in (\ref{eq:enh}) is not protected by Sudakov factors,
and the hope
is that at very low values of $x_4$ the saturation momentum $Q_s(x_4)$ is large
enough for pQCD to be applicable. However nothing is known about the gluon density in this region, where many other more complicated multi-Pomeron graphs, not
accounted for in the BK-equation, become important. This pQCD estimate is based
on the simplest Reggeon graphs, whose validity is questionable close to the saturation regime. The true parameter of the perturbative series is not the QCD coupling $\alpha_S$, but 
the probability of additional interactions, which however tends to 1 as the saturation region is approached.

There are several phenomenological reasons why the contributions of the enhanced diagrams are expected to be small\cite{KMRenh}.

\begin{itemize}
\item[(a)]
The NLL correction to $V_{3P}$ has the form of a rapidity threshold with $\Delta Y \simeq 2.3$; that is the rapidity between the $3P$-vertex and the proton, and between the $3P$ and the Higgs vertices, both must exceed $\Delta Y$. Thus there is pratically no phase space for such a contribution to produce a $m_H \sim 120$ GeV Higgs at the LHC.
\item[(b)]
If the enhanced diagrams were important, then they should also contribute to $\sigma_{\rm tot}$ and $\sigma^{\rm diffractive}_{\rm dissociation}$. Thus to calculate ${\hat S}^2$ we need to describe the ``soft'' data with enhanced rescattering included. This leads to a redistribution of the absorptive effects between the eikonal and enhanced contributions, but does not appreciably change the prediction for the total ${\hat S}^2$. Analogously, the prediction for $\sigma_{\rm tot}$ at LHC has a very weak model dependence if the model describes existing ``soft'' data. This is confirmed by the fact that $\sigma^{\rm diffractive}_{\rm dissociation}$, which is very sensitive to enhanced absorption, is almost flat in energy from about $\sqrt {s} \sim 50$ GeV. So we do not expect any extra suppression of diffraction at the LHC.
\item[(c)]
The leading neutron spectra at HERA, $\gamma p \to Xn$, are mediated by $\pi$-exchange.  Here the associated rapidity interval, $\Delta Y_{\pi}$, available for the $3P$-vertex, is large enough for an enhanced contribution.  Since $\Delta Y_{\pi}$ grows with the initial photon energy, an enhanced contribution would lead to a decrease of leading neutrons\cite{ln}.  This is not seen in the HERA data. 
\end{itemize}

Finally, valuable checks of the exclusive formalism can come from observing either exclusive $\gamma \gamma$ or dijet production at the Tevatron. For the former 3 events have been seen\cite{CDF}, in agreement with expectations\cite{KMRS}.  Dijet production is complicated by hadronization and detector resolution effects, which smear out the exclusive peak expected at $R_{jj} \equiv M_{jj}/M_X=1$.  Studies are underway to improve the identification of the exclusive signal, for instance by using a variable $R_j$ defined in terms of the largest $E_T$ jet\cite{Rj}.

\section{Acknowledgements}
We thank A.B. Kaidalov for valuable discussions.  ADM thanks
the Japan Society for the Promotion of Science for a Fellowship.

\end{document}